\documentclass[fleqn,twoside]{article}
\usepackage{espcrc2}

\usepackage{graphicx}
\usepackage[figuresright]{rotating}

\newcommand{\AmS}{{\protect\the\textfont2
  A\kern-.1667em\lower.5ex\hbox{M}\kern-.125emS}}

\title{Solar Activity during Two Millennia as Estimated from Annual Tree Rings}

\author{Y. Muraki\address{ Solar-Terrestrial Environment Laboratory, Nagoya University, 
        Nagoya 464-8601, Japan}%
        \thanks{email: muraki@stelab.nagoya-u.ac.jp},
  T. Mitsutani\address{Nara National Research Institute for Cultural Properties, 
         Nara 630-8577, Japan\\},
  S. Kuramata\address{Faculty of Science and Technology, Hirosaki University, 
         Hirosaki 036-8561, Japan\\},
  K. Masuda$^{a}$,
  K. Nagaya$^{a}$, 
  and
  S. Shibata\address{Engineering Science Laboratory, College of Engineering, Chubu University,
            Kasugai 487-0027, Japan\\}}

\begin{document}

\begin{abstract}
  The relationship between solar activity and the global climate is not only an academically interesting issue, but also an important problem for human beings. Lean and Rind have analyzed a considerable amount of climate data from around the world from 1889 to 2006. According to their analysis, the global effect was estimated to be 0.17{$\pm$}0.01 K between the solar maximum and minimum. However, they noticed that the effect strongly appeared in the zones between 
70{$^\circ$}N and 30{$^\circ$}N, and between 25{$^\circ$}S and 50{$^\circ$}S.
At its peak latitude (near 40{$^\circ$}), the effect was estimated to be 0.5 K.
Therefore, we analyzed a tree that survived at the Murooji temple 
in Nara Prefecture, Japan, for 391 years. 
Quite surprisingly, Fourier analysis of the annual growth rate identified two cycles 
with periodicities of 12 and 25 years during the Maunder minimum.
We have continued the analysis, using 
different samples from the Nagusa shrine in Hyogo Prefecture, Itayanagi City,
Aomori Prefecture, Japan, and from Yaku Island in Kyusyu, Japan.
An evidence of solar activity was found in all of the samples.
In particular, we have discovered a correlation
between Swiss glacier fluctuation and the growth rate of the Yaku
tree ring.
\vspace{1pc}
\end{abstract}

\maketitle

\section{Introduction}

  At the 2012 International Cosmic Ray Conference in Beijing \cite{bib:muraki1}, 
we presented evidence that annual tree rings provide a record of past solar
 activity \cite{bib:dengel}.  A cedar tree that survived for 
391 years at the Muroji temple, Nara Prefecture, Japan
(34{$^\circ$}32{$^\prime$}N, 136{$^\circ$}2{$^\prime$}E), indicated 
that the dynamo mechanism of the Sun occurred during the Maunder minimum (1645-1725).  
A Fourier analysis of the width of its annual tree rings showed 
a 12- and 25-year periodicity during 1606 and 1745, respectively.  
However, such a clean signal could not be found in the data set
of Muroji after the Maunder minimum.

  Possible reasons for this problem were presented in our previous paper. One reason can be related to the growth rate of trees.  In their early stages of life, trees grow rapidly, and while in their later stages, they grow at a more gradual rate.  
The width of the cedar tree at the Murooji temple only increased about 0.5 mm per year after 1950 (the cedar tree was uprooted by a typhoon in September 1998).  
Therefore, it could be difficult to identify such small differences in annual growth rate in a tree's outer rings.  
Furthermore, there is a possibility that the variation in climate after the Maunder minimum that was caused by solar activity was moderate in comparison with such a variation during the Maunder minimum \cite{bib:muraki2}.  
Therefore, we decided to seek further evidence from new samples and have analyzed the growth rate via the Fourier analysis method.  These results are reported in the current paper.
 
  We have also found a very valuable data set from the annual tree rings 
of a Yaku-cedar ({\it Cryptomeria japonica}) that recently survived in the southern 
area of Yaku Island, Japan, between 64 AD and 1988 AD.  In the current paper, 
we report the results of wavelet analysis on the tree rings of the Yaku-cedar, 
and compare the results with data of glacier fluctuation 
of the Swiss Alps \cite{bib:holzhauser}.

\section{Tree Ring Samples}

To confirm the previous analysis on the cedar tree collected from the Murooji temple, a tree ring that formed during the Maunder minimum (1645-1725) was analyzed.  The tree was collected from the Nagusa shrine in Hyogo Prefecture 
(35$^\circ$25{$^\prime$}N, 134$^\circ$40{$^\prime$}E). The shrine is located at an altitude of 750 m ASL and is in a region covered with heavy snow during the winter.
This sample is henceforth referred to as {\it Sample A}.

A young tree sample was also investigated.  It was collected from the town of Itayanagi near Hirosaki City of Aomori Prefecture. The coordinates of this location are
140{$^\circ$}27{$^\prime$}11 {$^\prime$$^\prime$}E and 
 40{$^\circ$}41{$^\prime$}57 {$^\prime$$^\prime$}N.
Itayanagi is located at 16.7 m ASL and is covered with snow during the winter.  The tree was a zelcova and was planted in 1987 and cut down in 2006. This tree sample is henceforth referred to as {\it Sample B}.

A cedar tree sample that was collected from Yaku Island was also investigated. The approximate coordinates of Yaku Island are 130$^\circ$30 {$^\prime$}E 
and 30$^\circ$25 {$^\prime$}N.  The island has a mountain that peaks at 1,936 m ASL; the cedar tree lived on the slope of this mountain.  This tree sample is henceforth referred to as {\it Sample C}.  From previous dendrochronology, the tree rings of this sample indicated that the tree lived between 64 AD and 1988 AD, which is nearly two millennia \cite{bib:mitsutani}.  This makes the sample very valuable in terms of searching for evidence on climate change due to solar variability.  However, it should be noted that for the present analysis, this sample's growth rate after 1000 AD is based on the average growth rate of several Yaku cedar trees, which survived at an altitude around 1,000 m ASL.

Table 1 shows the location each sample survived, period the tree survived, and the sample's tree type.  The original sample (O) from the Muroji temple is also listed in the table.
\begin{table}[h]
\begin{center}
\begin{tabular}{|l|c|c|}
\hline (Sample) Location  & Period (AD) & Tree type \\ \hline
(A)  Nagusa      & 1556-1992  & cedar\\ \hline
(B)  Itayanagi   & 1887-2006 & zelcova \\ \hline
(C)  Yaku        & 64-1988  & cedar\\ \hline
(O)  Muroji     & 1606-1998 & cedar \\ \hline
\end{tabular}
\caption{The tree samples used in the current analysis.}
\label{table_single}
\end{center}
\end{table}

\section{Data Analysis}
  For {\it Sample A}, we have compared the year-by-year growth rates. 
A good correlation between the sample A and B was found with a correlation factor of 0.74.  
Then, a three-year running average data set was created.  
A Fourier analysis of this data set was conducted, and 
as shown in Figure 1, 12-, 14- and 24-year periodicities were produced.

  For {\it Sample B}, a simple Fourier analysis was applied to the raw data.  The results are shown in Figure 2.  The results indicate that even for a modern tree ring sample, evidence of solar activity could be found with an 11-year periodicity.  This may confirm an earlier prediction by Dengel et al. \cite{bib:dengel}.  They analyzed modern pine trees from Scotland and found a relationship between solar activity and the width of a tree ring.

  For {\it Sample C}, a three-year running average data set was created.  Then, a Fourier analysis was applied to this data set for two different durations: between 200 and 1200 AD (data set I) and between 1200 and 1988 AD (data set II).  The data set was separated into two sections in order to clearly identify each periodicity.  Furthermore, a Fourier analysis was not applied to the earlier tree rings (64-200 AD) since the growth rate of the tree was very rapid.

  Figures 3 and 4 show the results for the analyses of data sets I and II, respectively.  For data set II, an 11-year periodicity and a 24-year periodicity can be clearly observed.  As shown in Figure 5, this tendency is clearer when the Fourier analysis was applied to the data set between 1360 and 1988 AD. These periodicities are believed to be induced by solar activity.

For data set I, periodicities of 10, 12.5, 13.5, 16, 20, 24, 27, 29, 32 and 38 years were observed.  The periodicities of 20, 24, 27, 29, 32 and 38 years may be possible higher overtones of the 10-, 12.5-, 13.5-, 16- and 20-year periodicities. Therefore, a wavelet analysis of the data set is needed. The results are presented in the following section.

\begin{figure}[t]
  \centering
  \includegraphics[width=0.45\textwidth]{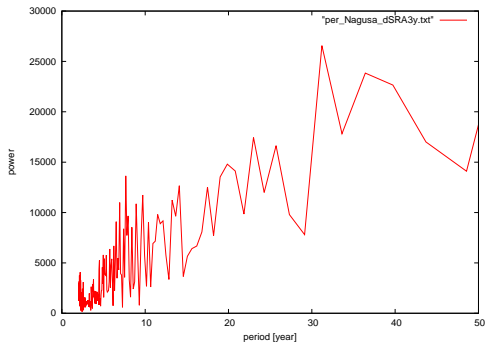}
  \caption{Results of the Fourier analysis for Sample A.}
  \label{simp_fig}
 \end{figure}
 
\begin{figure}[t]
  \centering
  \includegraphics[width=0.48\textwidth]{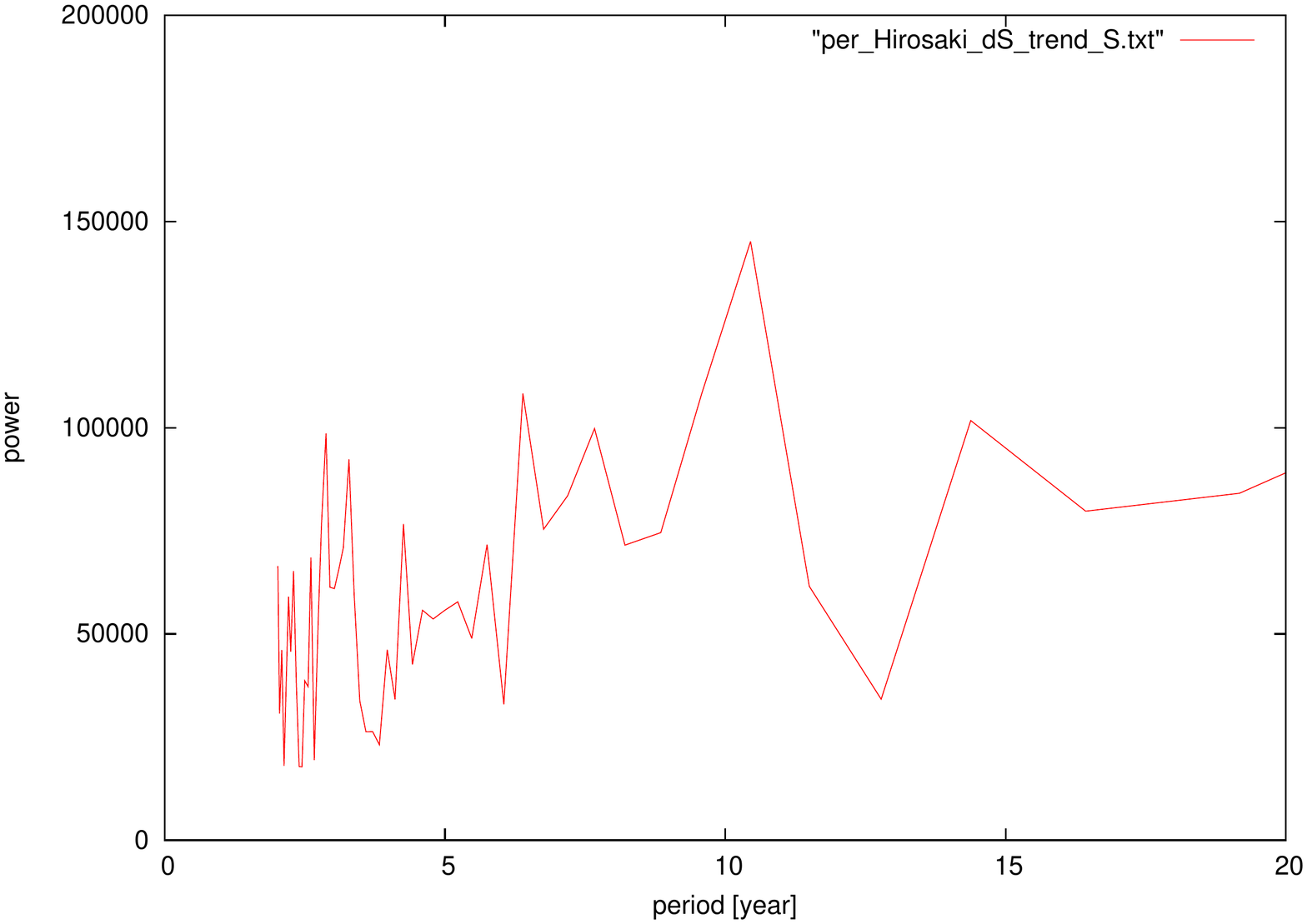}
  \caption{Results of the Fourier analysis for Sample B.}
  \label{simp_fig}
 \end{figure}
 
 \begin{figure}[t]
  \centering
  \includegraphics[width=0.45\textwidth]{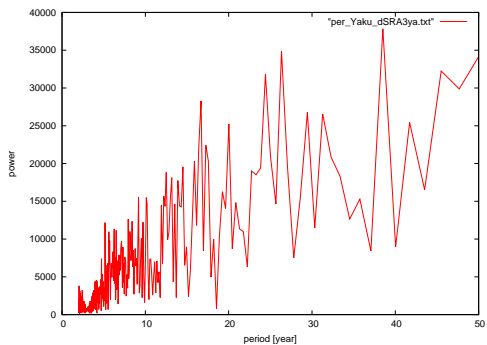}
  \caption{Results of the Fourier analysis for Sample C during 200-1200 AD (dataset I).}
  \label{simp_fig}
 \end{figure}
 
  \begin{figure}[t]
  \centering
  \includegraphics[width=0.45\textwidth]{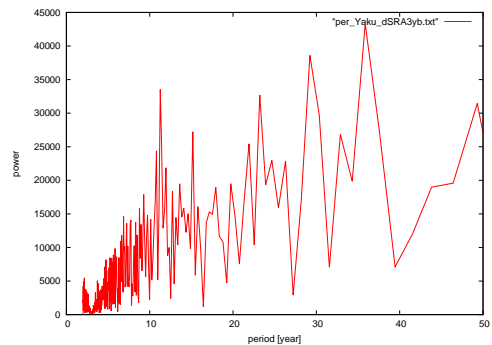}
  \caption{Results of the Fourier analysis for Sample C during 1200-1988 AD (data set II).}
  \label{simp_fig}
 \end{figure}
 
 \begin{figure}[t]
  \centering
  \includegraphics[width=0.48\textwidth]{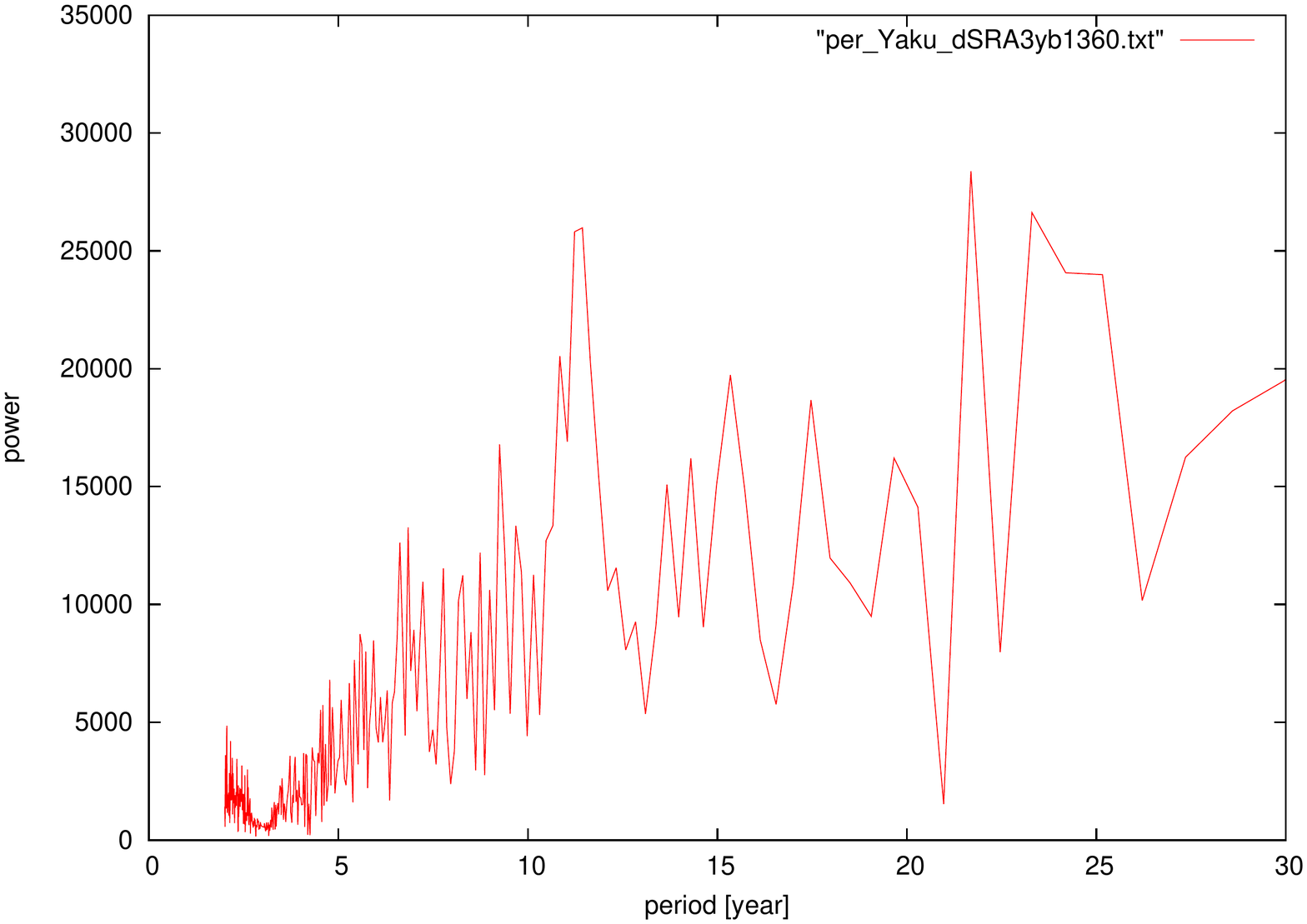}
  \caption{Results of the Fourier analysis for Sample C during 1360-1988 AD.}
  \label{simp_fig}
 \end{figure}

\section{Wavelet Analysis Results}

  The results of the wavelet analysis are shown in Figure 6.  An 11-year and a 22-year periodicity can be seen in the figure.  The 11-year periodicity is dominantly seen with the 22-year periodicity during 1100 and 1900 AD.  The vertical axis corresponds to the inverse of the periodicity (i.e., 0.1 corresponds to a 10-year periodicity).  To see this tendency more clearly, a wavelet analysis was applied to a 5-year running average data set instead of a 3-year running average data set.  The results are shown in Figure 7.

  If this figure is observed carefully, one can notice that the 11-year periodicity is not clearly recognized during earlier years.  We currently do not know if this is a natural occurrence or a sampling effect of the tree rings.  
However, several peaks are defined around 310, 410, 640, 840 AD during the period between 100-1000 AD.



\
 \begin{figure}[t]
  \centering
  \includegraphics[width=0.52\textwidth]{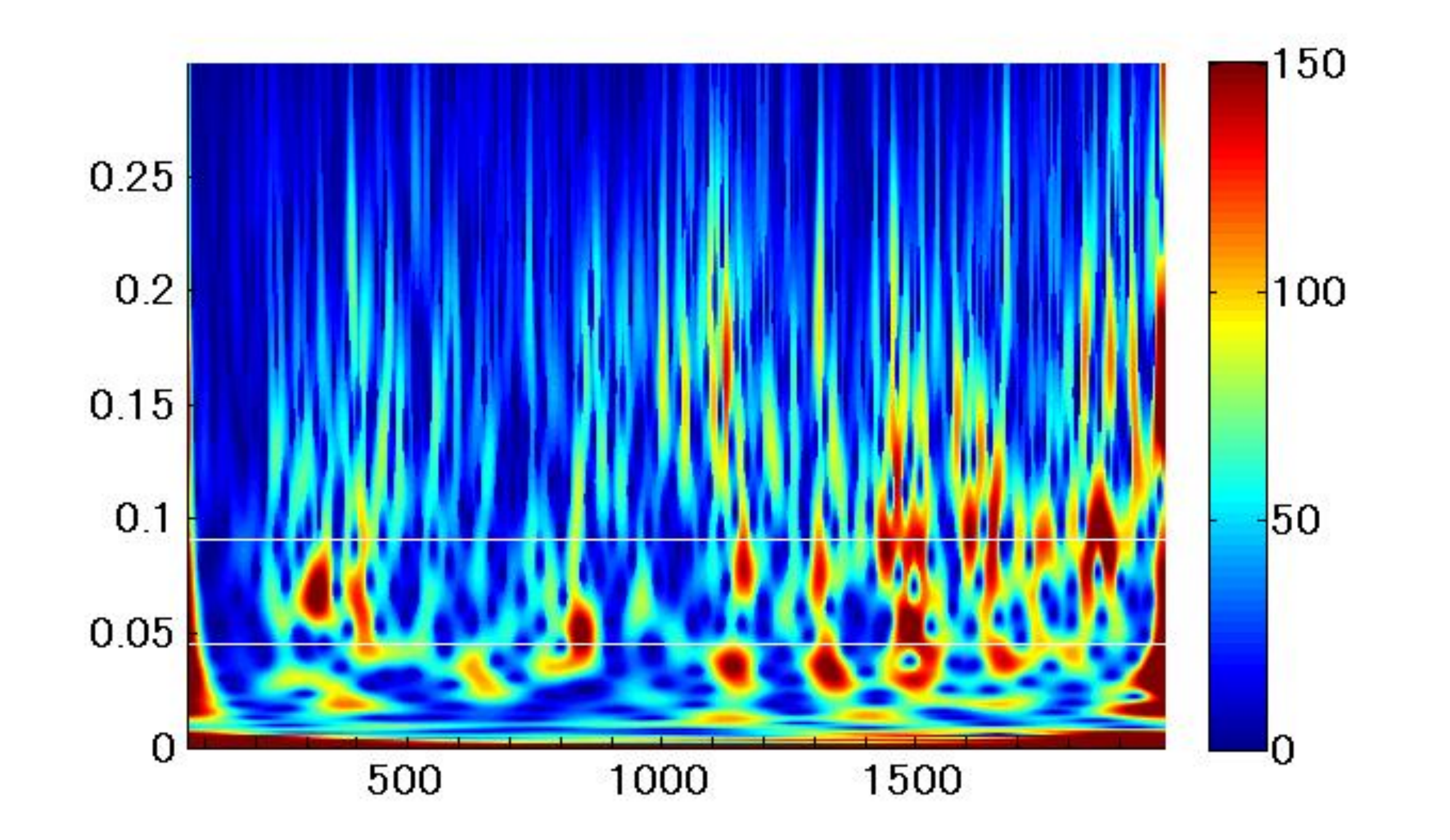}
  \caption{Results of the wavelet analysis for Sample C for the
   3-year running average data set.  The numbers under the horizontal axis 
   represents the year when the cedar tree survived (A.D.). }
  \label{simp_fig}
 \end{figure}
 
 \begin{figure*}[!t]
  \centering
  \includegraphics[width=1.10\textwidth]{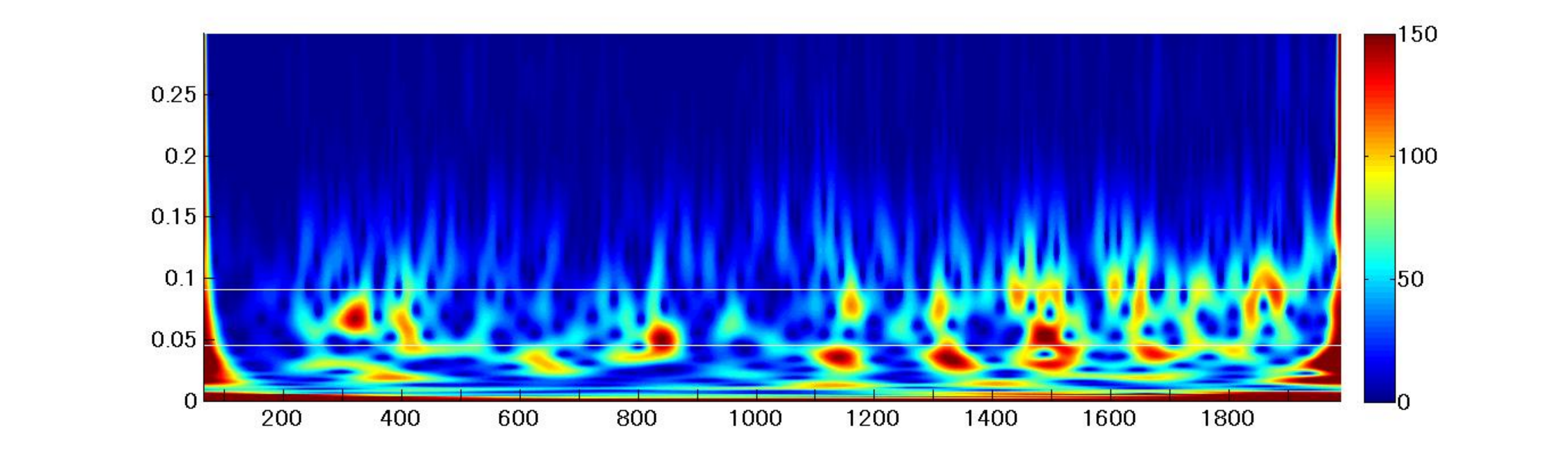}
  \caption{Results of the wavelet analysis for Sample C for the
   5-year running average data set.  Horizontal white lines indicate
   the inverse of the 11-year and 22-year periodicities.
   The numbers under the horizontal axis represents the year when the
   cedar tree survived. }
  \label{simp_fig}
 \end{figure*}
 
\section{Results and Discussions}

We now compare the present results with other results obtained 
via the measurement of glacier length.  Swiss scientists have been 
measuring the length of the glaciers 
of the Swiss Alps and a thorough report is now available on their historical record.  
We have compared our data with the data provided 
by H. Holzhauser, M. Magny, and H.J. Zumbuhl \cite{bib:holzhauser}.

A summary of this comparison is presented in Table 2.
The table states the dominant period when the 11- and 
22-year periodicities were recognized in the
tree rings, the longest year in which the Aletsch (Swiss) glacier was present, 
and additional comments.
It should be noted that during Grand minima,
a 22-year periodicity and an 11-year-periodicty are seen.

It is interesting to see that the advancing/retreating of the glacier is deeply correlated with the growth rate of Japanese cedar trees.  Our data were taken from locations within Japan, which are warmer regions when compared to the cool European mountains of which the glacier data were obtained. Thus, it is natural to believe that the past world climate consisted of low temperatures for both Japan and Europe.

However, our Fourier analysis and wavelet analysis indicate an 11-year periodicity and
a 22-year periodicity.  This implies that this phenomenon was induced by {\it solar activity}.

Although, there are some differences between our data and the glacier data.  It is slightly unclear why the glacier data do not include a possible extension of the glacier during the Spoerer minimum. Our data suggests such an extension during 840 AD.  This may be related to the occurrence of the 14- to 22-year periodicity during 200 and 900 AD, which was most likely due to a meteorological effect of the global climate \cite{bib:beer}.

In fact, the 11-year periodicity was found in the last excess spot around 1860.  However, it does {\it not} associate well with the 22-year periodicity.  Therefore, the last spot found in the tree ring may have been produced by something different.


Furthermore, we compared our data with that obtained via the measurement of heavy water (O18) in the Yaku cedar tree \cite{bib:kitagawa}.  However, we could not find any strong correlation between both data sets (data not shown).

\begin{table}[h]
\begin{center}
\begin{tabular}{|l|c|c|}
\hline Period (tree) &  Year (glacier)  &  Comments \\ \hline
1860 {$\pm$} 40  & 1859/60 & no 22-year  \\ \hline
1670 {$\pm$} 40  & 1669/70 & Maunder \\ \hline
1490 {$\pm$} 40  &  ----?  & Spoerer \\ \hline
1330 {$\pm$} 40  & 1369    & Wolf    \\ \hline
1145 {$\pm$} 40  & 1160    & Oort       \\ \hline
 840 {$\pm$} 30  &  850    & no 11-year \\ \hline
 640 {$\pm$} 40  &  650    & no 11-year \\ \hline
 410 {$\pm$} 20  & ----    & 14-year?   \\ \hline 
 310 {$\pm$} 40  &  300    & 14-year?   \\ \hline
\end{tabular}
\caption{Comparison between the longest year the Aletsch glacier was present 
and the peak year of the 11- and/or 22-year periodicities 
were observed in accordance with the tree rings. }
\end{center}
\end{table}

 \begin{figure}[t]
  \centering
  \includegraphics[width=0.50\textwidth]{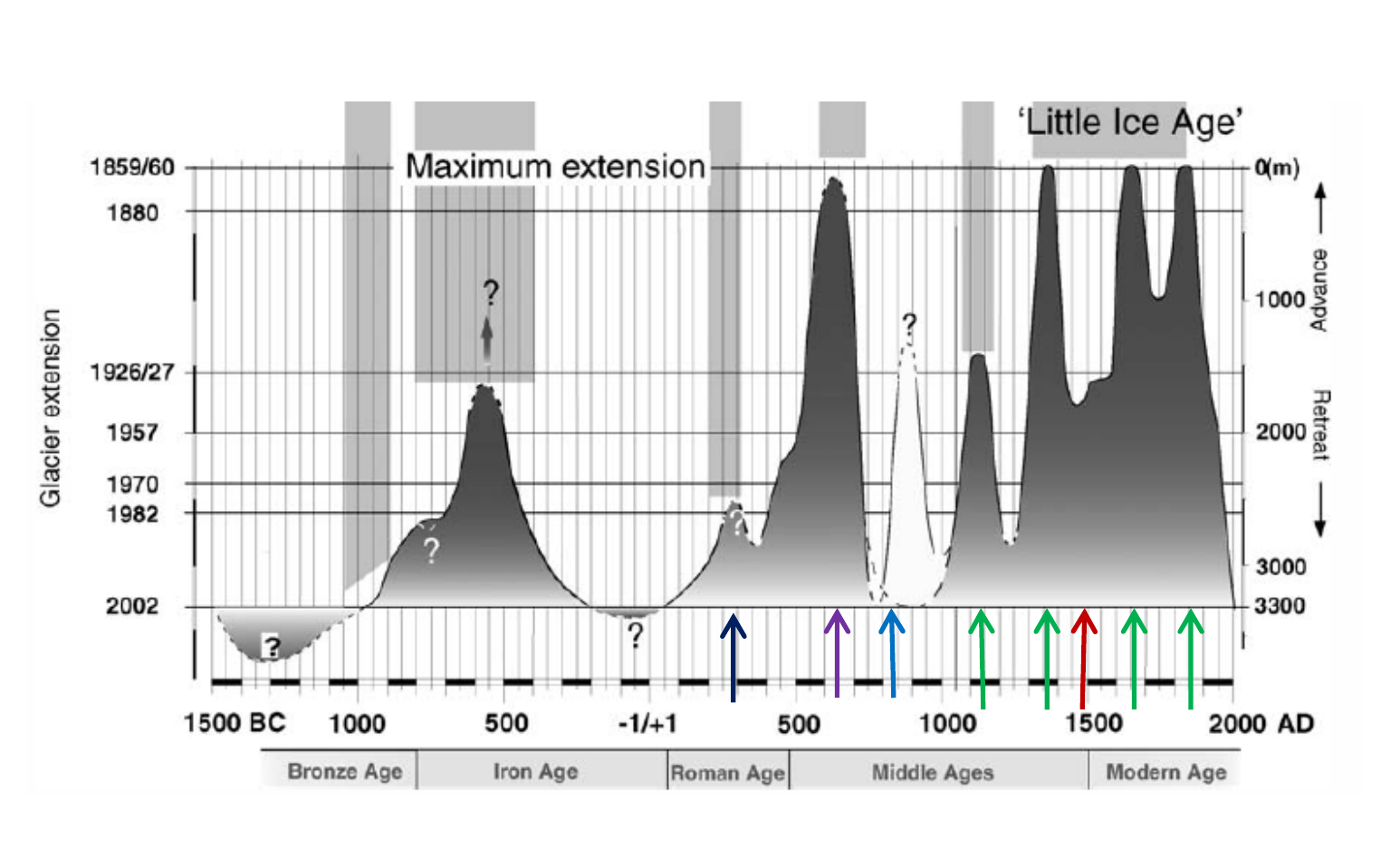}
  \caption{The Great Altsch glacier extension curve of Switzerland
    [6]. The green arrows correspond to the strong 
    24-year periodicity periods of the tree rings.
   The brown arrow indicates strong evidence 
    of the glacier advancing during the Spoerer minimum.  
The purple and blue arrows correspond 
   to the peak year for periodicity of the tree ring, 
but only a 22- to 24-year periodicity is recognized.  
The black arrows corresponds to the 14-year periodicity.  
It should be noted that the growth rates of the tree ring of this duration
(before 10th century) have only been measured by a single large tree.}
  \label{simp_fig}
 \end{figure}
 
\section{Conclusions}

We conducted our research according to the prediction made by Lean and Rind \cite{bib:lean}.  Some indications of solar variability were found in Japanese cedar tree rings.  In particular, such an effect was dominantly seen during the grand minima, initiating at the start of the 12th century and concluding at the end of 18th century.  The effect of solar radiation was not permanent, but it did affect the global climate.  Particularly, the occurrence of solar radiation during the grand minima suggests that the dynamo mechanism of the Sun also occurred.  Therefore, we predict that an enhancement of approximately 1145 will be produced via solar variation.

\vspace*{0.5cm}
\footnotesize{{\bf Acknowledgment: }{The authors would like to acknowledge those who kindly provided the tree samples.  T. M. acknowledges the Yaku Forest Office, which provided him with very valuable tree samples.  S. K. and Y. M. thank Mr. Tsubota and the town of Itayanagi for providing them with a tree sample.}


\begin{thebibliography}{}

\bibitem{bib:muraki1} Y. Muraki, K. Masuda, K. Nagaya, and K. Wada, Proceed. of 32nd ICRC (Beijing), 11 (2011) 424-427. doi:10.7529/ICRC2011/V11/0369.

\bibitem{bib:dengel} S. Dengel, D. Aeby, and J. Grace, New Phytologist, 184 (2009) 545-551. doi:10.1111/j.1469-8137.2009.03026.x.

\bibitem{bib:muraki2} Y. Muraki, K. Masuda, K. Nagaya, K. Wada and H. Miyahara, Astrophysics and Space Science Transactions, 7 (2011) 395-401. doi:10.5194/astra-7-395-2011.

\bibitem{bib:mitsutani} T. Mitsutani, The data of the tree rings
were obtained from the Yaku Forest Office by the use of a measuring machine. 
\bibitem{bib:lean} J. L. Lean and D. H. Rind, Geophysical Research Letter, 35 (2008) L18701. doi: .

\bibitem{bib:holzhauser} H. Holzhauser, M. Magny and H. J. Zumbuhl, The Holocene, 15 (2005) 789-801. doi: .

\bibitem{bib:beer} J. Beer, Space Science Review, 11 (2000) 53-66.
\bibitem{bib:kitagawa} H. Kitagawa and E. Matsumoto, GRL 22 (1995) 2155-2158. doi:0094-8534/95/95GL-02066.


\end{thebibliography}
\end{document}